\documentclass[aps,prl,twocolumn,groupedaddress]{revtex4-1}
\pdfoutput=1

\usepackage{amsmath}
\usepackage{amssymb}
\usepackage{lastpage} 
\usepackage{multirow}
\usepackage{graphicx}
\usepackage{float}
\usepackage{midfloat}

\begin{document}

\title{Spectral and temporal characterization of a fused-quartz microresonator optical frequency comb}

\begin{abstract}
We report on the fabrication of high-$Q$, fused-quartz microresonators and the parametric generation of a frequency comb with 36 GHz line spacing using them.  We have characterized the intrinsic stability of the comb in both the time and frequency domains to assess its suitability for future precision metrology applications.  Intensity autocorrelation measurements and line-by-line comb control reveal near-transform-limited picosecond pulse trains that are associated with good relative phase and amplitude stability of the comb lines.  The comb's 36 GHz line spacing can be readily photodetected, which enables measurements of its intrinsic and absolute phase fluctuations.
\end{abstract}

\author{Scott B. Papp and Scott A. Diddams}
\affiliation{National Institute of Standards and Technology, Boulder, Colorado 80305, USA}

\date{\today}
\maketitle

Femtosecond-laser optical frequency combs have revolutionized frequency metrology and precision timekeeping by providing a dense set of absolute reference lines spanning more than an octave.  These sources exhibit sub-femtosecond timing jitter corresponding to an ultralow phase noise of $<100$ $\mu$rad on high harmonics of their typically 100's of MHz repetition frequency (line spacing) \cite{Diddams2004}.  This remarkable level of performance has enabled measurements of atomic transition frequencies at the 17th digit \cite{Rosenband2008} and the generation of ultralow noise signals \cite{Fortier2011}.   Beyond their natural application as an optical clockwork, frequency combs are used in diverse applications including precision measurements of fundamental physics \cite{Diddams2004,Rosenband2008}, direct spectroscopy and real-time trace gas detection \cite{Thorpe2006}, molecular fingerprinting \cite{Diddams2007}, astronomical spectrograph calibration \cite{Murphy2007}, and optical arbitrary waveform generation\cite{Huang2008}.  An even broader range of applications may be possible if the bulk and complexity of a femtosecond comb could be reduced without significantly increasing fluctuations in the comb spectrum.

A new class of frequency combs has recently emerged based on monolithic microresonators, henceforth denoted microcombs \cite{DelHaye2007,Savchenkov2008,Levy2010,Razzari2010,Kippenberg2011}.  Here the comb generation relies on parametric conversion provided simply by third-order nonlinear optical effects and is enabled by recent advances in the quality factor $Q$ and the small volume of microresonators.  These devices require only a single continuous-wave laser source, but the usable frequency span of the comb depends on low dispersion, making material properties critical. Microcombs also present a unique platform for creating large line spacings (10's of GHz to THz).  To date microcombs possessing 100's of lines each spaced by 100's of GHz have been created with silica microtoroids \cite{DelHaye2007} and silicon-nitride microrings \cite{Levy2010,Razzari2010}.  Uniformity \cite{DelHaye2007} and control \cite{DelHaye2008} of silica microcombs have also been demonstrated.  Microcombs with mode spacings of 10 to 25 GHz have also been realized with machined crystalline resonators \cite{Savchenkov2008} and aspects of their microwave-domain spectral purities have been analyzed \cite{Savchenkov2008,Savchenkov2008a}.

\begin{figure}[t]
\centering
\includegraphics[width=0.45\textwidth]{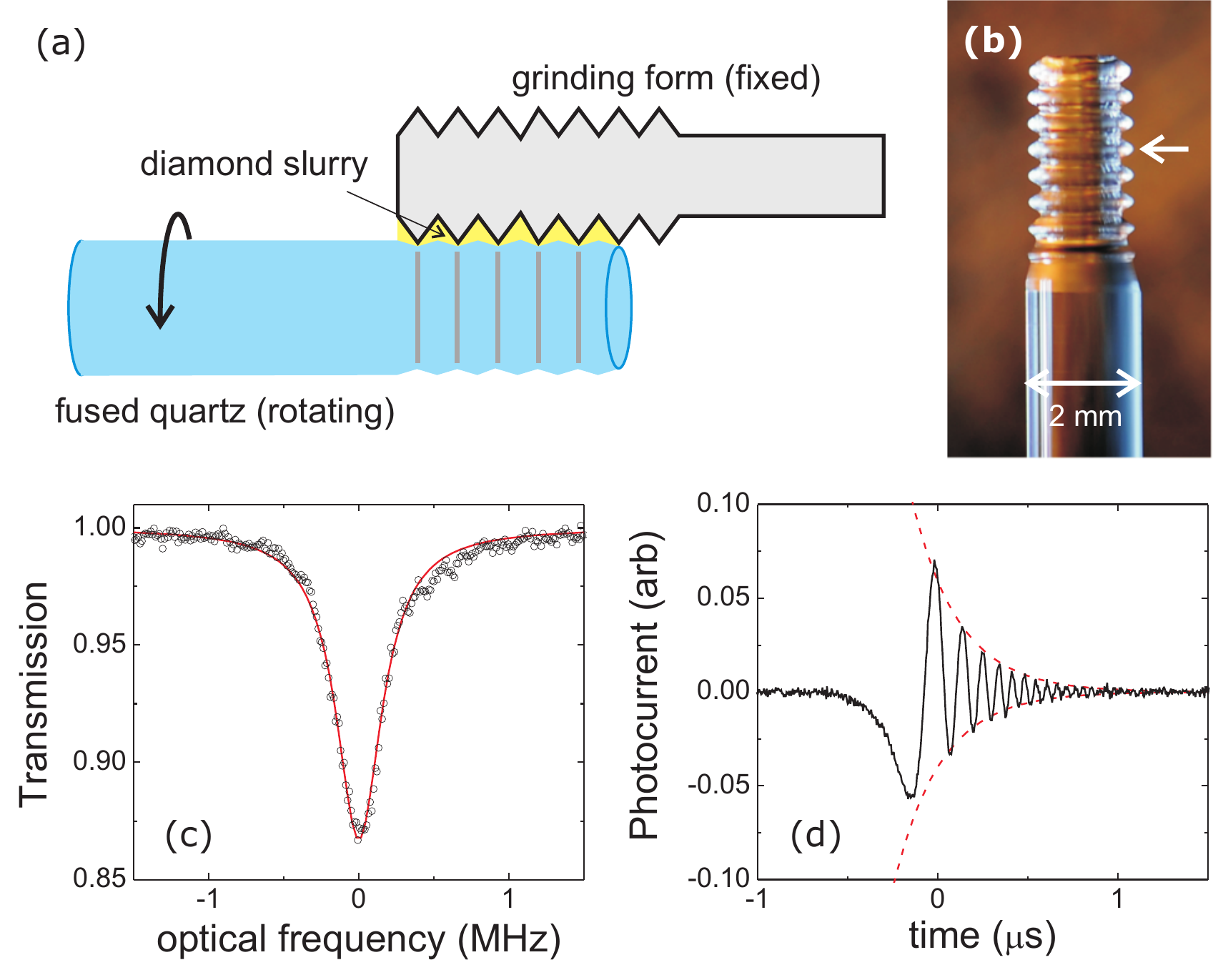}
\caption{High-$Q$ fused-quartz microresonators. (a)  A fused quartz rod is rotated in a lathe (not shown) and rubbed against a fixed metal form charged with diamond abrasive.  The ground form is then flame polished with household-grade propane and oxygen (not shown). (b)  Image showing an array of fused quartz microresonators after grinding and flame polishing.  The disks of 1.9 mm diameter are separated by 0.45 mm. (c)  Frequency scan of a mode for disk four (indicated by the arrow) demonstrating an optical $Q_0 = 5.2\times10^8$.    (d)  Cavity ringdown at $K=1$ to determine the optical $Q$ of disk four.  The dashed lines indicate a decay time of 260 ns, corresponding to $Q_0 = 6.2\times 10^8$.
\label{fig1}}
\end{figure}

Future frequency metrology applications of microcombs will require a line spacing (repetition rate, $f_{\rm{rep}}$) in the 10's of GHz (millimeter-scale resonator), comb span approaching an octave, and low absolute phase and frequency fluctuations that can be further reduced by wideband comb control \cite{Kippenberg2011}.  Additionally, a threshold power for comb generation in the mW range, and the capability for integration with chip-based photonic circuits would enable novel portable applications.  Here we present a platform towards attaining these requirements.  Our fabrication approach takes advantage of precision mechanical shaping and polishing possible with fused quartz, a versatile amorphous material with low optical losses throughout the visible and into the infrared.  We created resonators that feature high $Q$ and low effective optical mode area, which allowed comb generation for a $<$10 mW pump laser.  By pumping above threshold, a comb with $\sim150$ lines spaced by 36 GHz was generated.  We harnessed full line-by-line control (amplitude and phase) of the comb to generate near-transform limited optical waveforms and to understand the comb's relative phase stability.  In the frequency domain, we characterized the comb's mode-spacing fluctuations to assess prospects for stabilizing it to optical atomic standards.

\begin{figure}[t]
\centering
\includegraphics[width=0.47\textwidth]{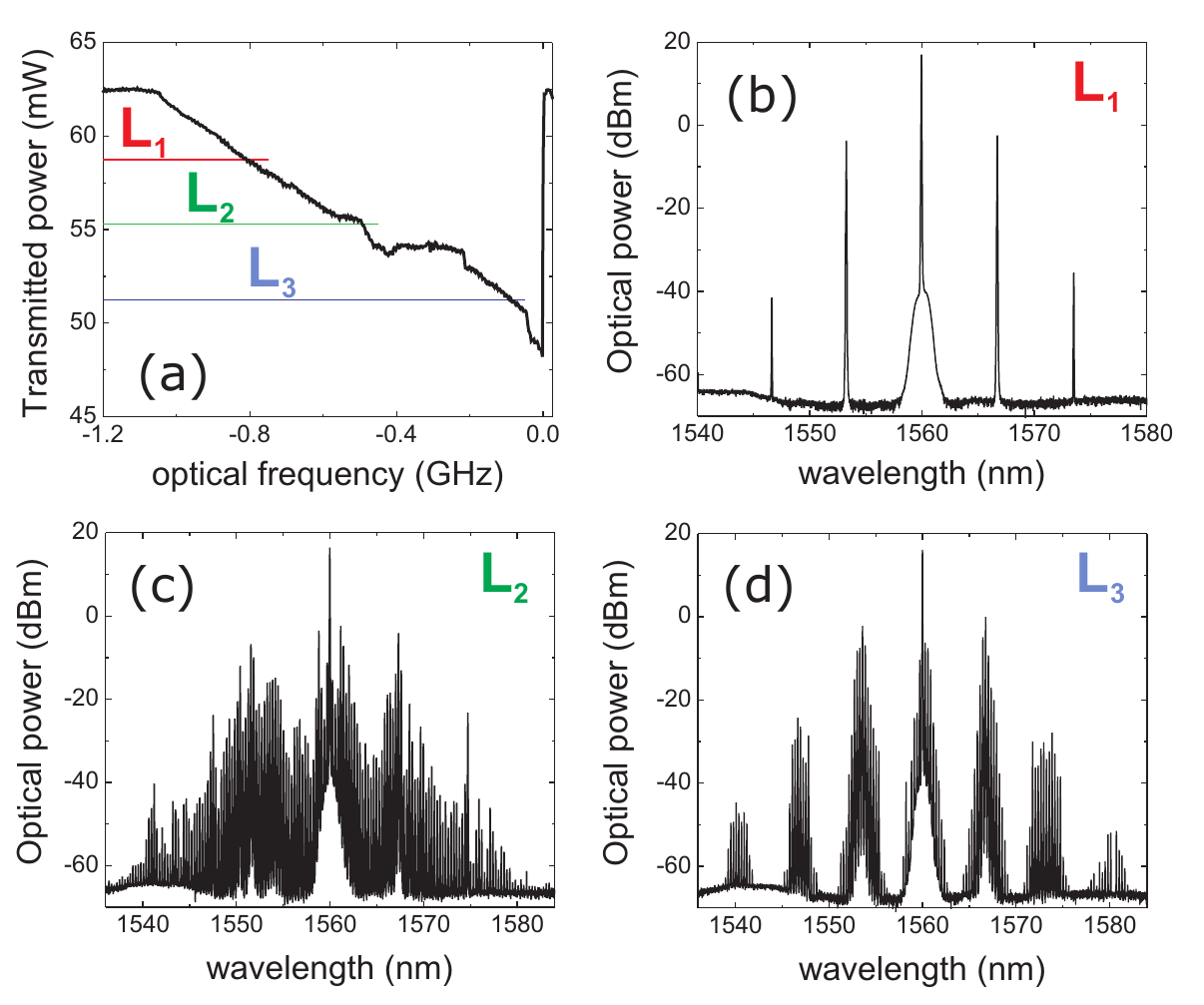}
\caption{Microresonator lineshape and corresponding comb spectra for laser pump power of 62 mW. (a)  Thermally broadened lineshape with a mostly sawtooth shape.  Power levels L$_1$, L$_2$, L$_3$ respectively correspond to the parametric threshold, the initiation of a 36 GHz comb but with unstable behavior, and the region of stable 36 GHz comb generation. (b)  Comb spectrum at L$_1$. (c)  Comb spectrum at L$_2$. (d)  Comb spectrum at L$_3$.
\label{lineshapes}}
\end{figure}

We fabricated an array of disk-like microresonators on a common substrate by use of a combination of diamond grinding and flame polishing.  A schematic of our procedure is shown in Fig. \ref{fig1}a.  Fused quartz rods with $\approx 2$ mm diameter were rotated in a ball-bearing spindle.  To create the basic shape of a disk, the glass was rubbed against a metal form with triangular cross section.  Diamond abrasive between them removed material from both.  At a diamond abrasive size of 6 $\mu$m, $\sim$30 minutes of grinding was required to create the underlying shape shown in Fig. \ref{fig1}b.  The resulting surface was far too rough to support high $Q$ whispering-gallery modes.  By polishing the disks with a propane-oxygen flame we achieved an extremely smooth surface.  Substantial melting of the quartz was required to alleviate subsurface damage caused by grinding.  Figure \ref{fig1}b shows an image of a completed disk array.  Free-spectral range measurements of several disks revealed a diameter uniformity of 0.1 $\%$.

After processing, the fused quartz rod was placed in a temperature-stabilized mount ($<$0.01 K stability) and whispering-gallery modes of the disks were accessed using standard fiber tapers \cite{Knight1997,Cai2000,Spillane2003}.  The coupling between disk modes and the taper can be characterized by the parameter $K = \tau_0 / \tau_e$, where $\tau_0$ is the intrinsic photon lifetime and $\tau_e$ is the photon loss rate due to coupling.  The frequency linewidth of the resonator-taper system is $\gamma = \tau_0^{-1} + \tau_e^{-1}$.  Figure \ref{fig1}c shows a frequency scan over a disk mode using a tunable diode laser at 1560 nm.  A low optical power of $\sim$10 $\mu$W was used to minimize thermal broadening of the resonance \cite{Carmon2004} and taper coupling was small ($K\approx0.03$).  The linewidth of the Lorentzian transmission feature is 0.37 MHz, which corresponds to an optical $Q_0 = \omega\,\tau_0$ of $5.2\times10^8$, where $\omega$ is the optical carrier frequency \footnote{We have not observed any scattering-induced mode splitting in these devices \cite{Gorodetsky2000,Kippenberg2002}.}.  We confirmed this $Q_0$ measurement by recording the resonator energy ringdown time following a rapid frequency scan across resonance with $K=1$; see Fig. \ref{fig1}d.  A fit to the cavity energy decay yields $Q_0=6.2\times10^8$, in good agreement with our linewidth measurement.  Material absorption in the fused-quartz samples we use imposes a limit to $Q_0 \sim 2 \times 10^{10}$.  We speculate that some contamination occurs during the flame-polishing step \cite{Kuzuu2003,*Kuzuu2004}.

To achieve the long-term stability in $K$ needed for studies of microcomb generation,  we place the taper directly in contact with the resonator surface.  Translating the taper along the axis of the quartz rod allows us to tune the resonator-taper coupling from under- to over-coupling.  Moreover, measurements of $K$ at various locations along the disk edge provide a qualitative picture of the mode's spatial profile.  Specifically, the resonator mode studied in Fig. \ref{fig1} c and d has two field nodes in the axial direction and is presumably of lowest radial order.  Numerical calculations based on an ideal disk model \cite{Borselli2005} suggest the effective area of this mode ($A_{\rm{eff}}$) is approximately 500 $\mu$m$^2$.  The threshold for parametric oscillation threshold is 
$P_{\rm{th}} = \frac{1}{4} \, \frac{1+K}{K} \, \frac{n}{n_2} \, \frac{\omega}{\Delta \nu} \, \frac{A_{\rm{eff}}}{Q^2}$, where $n$ ($n_2$) is the refractive index (nonlinear index) of fused quartz and $\Delta \nu$ is the free spectral range.  For our disk resonators we expect $P_{\rm{th}} = $ 4 mW for $K=0.2$.

We have generated a microcomb at, and significantly above, parametric threshold.  Figure \ref{lineshapes}a-d shows a disk resonator transmission lineshape and typical comb optical spectra.  With constant laser power, we adjusted and stabilized the intracavity power by tuning the pump laser frequency for a fixed $K=0.2$.  Here the lineshape is thermally broadened and assumes a characteristic ``sawtooth'' shape \cite{Carmon2004}.    Power level L$_1$ indicates the observed 3.7 mW threshold for parametric oscillation, which is in excellent agreement with our prediction.  Notably, the signal and idler modes are spaced by 828 GHz ($23 \times 36$ GHz); indicating the point at which resonator-induced phase mismatch of pump, signal, and idler balances the mismatch created by nonlinear effects \cite{Agrawal2007,Kippenberg2004b,Savchenkov2008}.  Increasing the intracavity power further resulted in a widened comb span up to $\pm$25 nm about 1560 nm and a reduction of the comb spacing toward the fundamental 36 GHz of the disk resonator, which is consistent with the simulations of Chembo \textit{et al.} \cite{Chembo2010,Chembo2010a}.  However, our observations indicate that 36 GHz comb generation is complex with competition between different modes of operation.  Specifically, we present two instances of 36 GHz combs with dramatically different properties: (1) At power level L$_2$ (Fig. \ref{lineshapes}c) an $\approx$36 GHz comb spacing is initiated, but a finer analysis (see Fig. \ref{abs}c) reveals numerous comb line spacing frequencies and (2) in a range about power level L$_3$ (Fig. \ref{lineshapes}d) a robust comb pattern exists with a single line spacing of 36.0012 GHz.  Interestingly, when we tuned to power level L$_2$ a deviation in the sawtooth shape of the transmission lineshape was observed.  This behavior depends on resonator-taper coupling and laser polarization, but not the laser frequency scan rate, which we varied from 0.005 to 1 s.  Moreover, in this range stable time-domain waveforms and low noise frequency-domain performance were not evident under the conditions of Fig. \ref{lineshapes} \footnote{The numerical simulation of parametric comb generation by Chembo appears to explain the spectrum of Fig. \ref{lineshapes}d with 36 GHz lines modulated at an approximately 828 GHz period, but not the sawtooth deviation at power level L$_2$ or the associated instability of comb generation here \cite{Chembo2010,Chembo2010a}}.  About L$_3$ we observed the resumption of a sawtooth lineshape.

\begin{figure}[t!]
\centering
\includegraphics[width=0.47\textwidth]{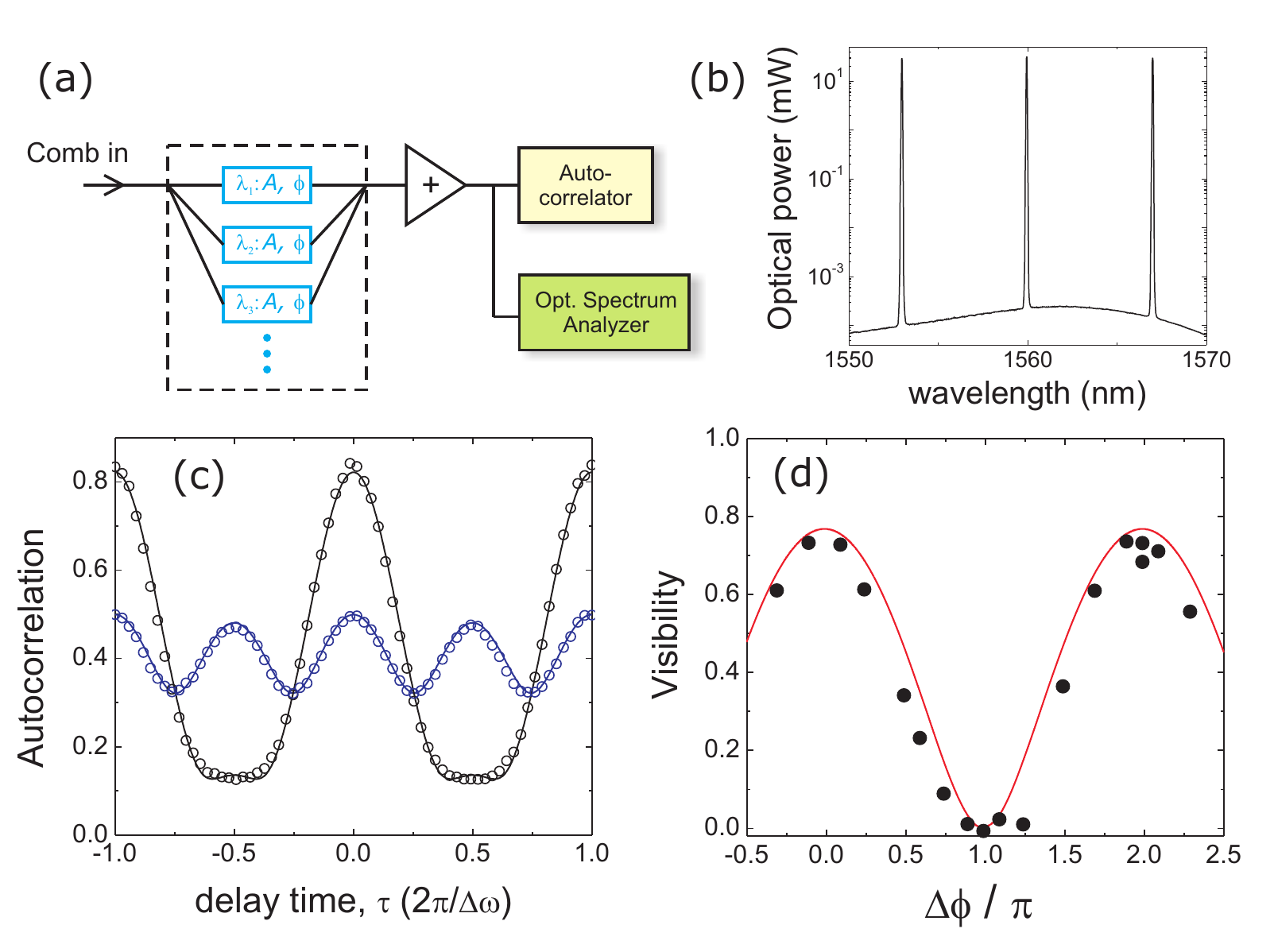}
\caption{Time-domain measurements of a three line comb. (a)  Schematic of our setup to control and analyze the microcomb spectrum.  The amplitude and phase of light exciting the resonator via tapered fiber is manipulated with a programmable filter.  An intensity autocorrelator is used to record the fields' waveforms. (b)  Optical spectrum following the filter and amplifier. (c)  Time-averaged autocorrelation signal for $\Delta\phi=0$ (black circles) and $\Delta\phi=\pi$ (blue circles), which respectively demonstrate constructive and destructive interference of the three comb fields.  The solid lines show the results of an analytic model. (d)  Visibility of constructive interference as a function of $\Delta \phi$.
\label{time1}}
\end{figure}

The spectral structure similar to Fig. \ref{lineshapes} has been observed in a variety of parametric microcombs, but little is known about the temporal structure of the output as determined by the phase relationship between the comb elements.  Measurements and control of optical waveforms can reveal crucial information about the internal fields of the microcomb \cite{Foster2011a,Ferdous2011a}. Using time-domain diagnostics, we studied the relative phase stability of the first three comb lines (pump, signal, and idler) at parametric threshold, and the phase stability of many lines far above threshold.  For a comb composed of only three equal-amplitude lines its time waveform has two distinct behaviors, which depend primarily on the relative phases of the lines: constructive interference, resulting in a rudimentary pulse at $f_{\rm{rep}}$, or partial destructive interference, resulting in a beat signal at $2 f_{\rm{rep}}$.  A comb with many lines exhibits more complicated behavior and offers high peak power waveforms.

\begin{figure}[b!]
\centering
\includegraphics[width=0.47\textwidth]{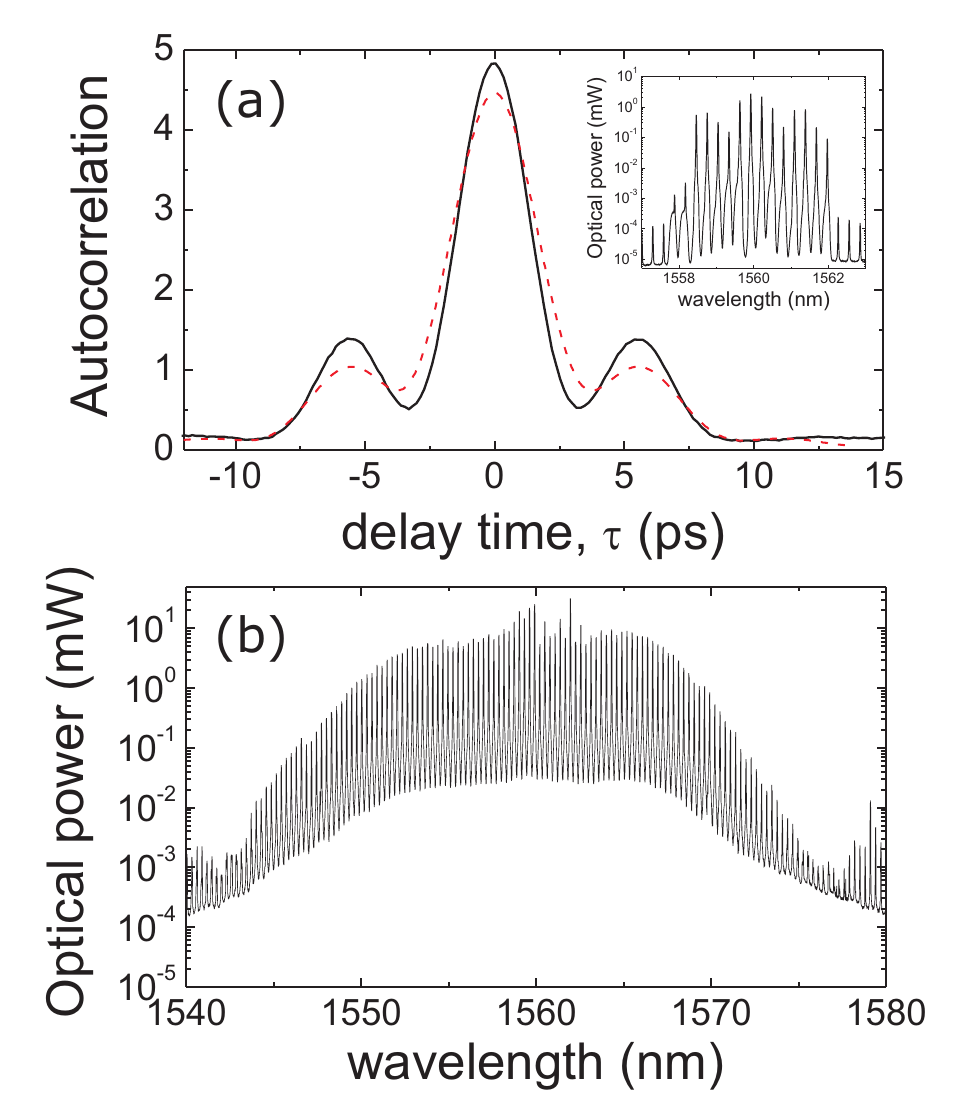}
\caption{Generation of picosecond pulses utilizing line-by-line comb control. (a)  Measured time-domain waveform (black line) following amplitude and phase optimization, and simulation (red dashed line) based on the measured amplitudes of comb lines used and assuming phase alignment of all the fields.  The inset shows the optical spectrum after optimization. (b)  Optical spectrum following broadening in highly nonlinear fiber.  The comb was amplified to 1.6 W.
\label{time2}}
\end{figure}

Our experimental starting point was a comb spectrum similar to that shown in Fig. \ref{lineshapes}b, except with spacing of 864 GHz  \footnote{The mode spacing at threshold can be tuned over a small range near 864 GHz with $K$.}.  Using a programable amplitude and phase filter featuring 10 GHz resolution (Fig. \ref{time1}a), we selected the central three lines of this comb, equalized their amplitudes to better than 7 \%, and adjusted their phases.  The optical spectrum we obtained following amplification with a low-dispersion erbium fiber amplifier is shown in Fig. \ref{time1}b.  It was delivered via a short section of optical fiber to an intensity autocorrelator for characterization.  By varying the relative phase ($\Delta \phi$) of the lowest wavelength line, we were able to explore the transition between constructive and deconstructive interference of the comb lines.  The relative phases between the center and the highest wavelength line were set to zero.  Figure \ref{time1}c shows the time-waveform autocorrelation signal ($P(\tau)$), where $\tau$ is the differential path delay, for $\Delta\phi=0$ (black open circles) and $\Delta\phi=\pi$ (blue open circles).  The solid lines indicate the results of an analytic model based on three phase-coherent fields, which are in excellent agreement with the data.  To characterize the interference between the three fields we introduce a visibility $V$ based on $P(\tau)$, which is defined as $V = |\frac{P(\pi/\Delta\omega) - P(0)}{P(\pi/\Delta\omega)+P(0)}|$.  The filled points in Fig. \ref{time1}d are measurements of $V$ as a function of $\Delta \phi$, and the red line shows our analytic model.  We measured a maximum (minimum) visibility of 0.7 (0.007), and the model predicts the visibility is 0.768 (0.002).  This indicates good phase coherence of the comb lines.  Fitting the model to the data with free parameter $\Delta \phi$ yields an uncertainty in $\Delta \phi$ of 20 mrad, which we associate with an upper limit on the fluctuations of the relative phase of the comb elements over the measurement time of 1700 seconds \footnote{We note this value also includes un-characterized fluctuations intrinsic to the programable filter and the measurement apparatus.}.

We also carried out time-domain waveform generation with a microcomb far above parametric threshold.  The amplitude and phase filter was used to adjust the central 15 lines (0.54 THz) of a 36 GHz comb similar to that of Fig. \ref{lineshapes}d.  Here a manual line-by-line procedure was used to optimize the peak waveform intensity.  This resulted in stable near-transform-limited 2.5 ps optical pulses, as demonstrated by the time-averaged autocorrelation signal in Fig. \ref{time2}a.  Hence, the relative phases of the 15 comb lines were presumed to be constant across the spectrum and were correspondingly stationary in time.  Moreover, the autocorrelation signal remained unchanged despite large perturbations to the microcomb system, such as re-locking the pump laser frequency or re-placing the tapered fiber in contact with the disk.  These observations indicate that the internal fields of the resonator and the pulse generation mechanism are deterministic and repeatable.  Significantly, the creation of time-domain waveforms, such as shown in Fig. 4a, implies that such a compact and simple device could be competitive with more conventional mode-locked lasers a source of high repetition rate ultrashort optical pulses that would be useful for a variety of time- and frequency-domain applications \cite{Kippenberg2011}. For example, nonlinear broadening in fibers could be a route to ultra-broadband spectra for comb self-referencing and comb spectroscopy.  Using a 101 m length of highly nonlinear fiber, we have broadened our 15 line comb by more than a factor of 10, as shown in Fig. \ref{time2}b. This demonstrates sufficient peak power to drive nonlinear processes external to the microresonator.

\begin{figure}[t]
\centering
\includegraphics[width=0.47\textwidth]{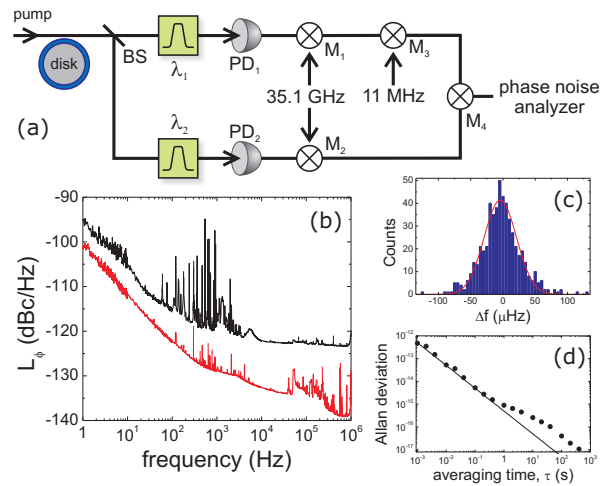}
\caption{Residual phase noise of microcomb line spacing. (a)  Schematic of our experimental setup.  The microcomb spectrum is split (BS) into two paths, which are separately filtered at $\lambda_1=1560$ nm and $\lambda_2=1553.6$ nm.  Photodectors PD$_{1,2}$ convert the comb spacing of the independent paths into electrical signals near 36 GHz.  PD$_1$ (PD$_2$) generates 2.15 mA (0.04 mA) of photocurrent.  Double-balanced mixers M$_1$ and M$_2$ convert these signals to baseband, and mixer M$_3$ offsets one path by 11 MHz.  This offset frequency was referenced to a hydrogen maser.  At mixer $M_4$ the two paths are interfered yielding residual noise.  (b)  Single-sideband phase noise of the microcomb 36 GHz tone (black line) and the contribution from the 11 MHz offset oscillator (red line).  (c)  Histogram of zero-dead-time counter measurements of the signal exiting M$_4$.   (d)  Allan deviation of the microcomb line spacing (filled points).  The solid line shows the $1/\tau$ scaling exhibited at most averaging times.
\label{rel}}
\end{figure}

Frequency-domain techniques offer significant advantages in characterizing microcomb fluctuations, including access to the power spectral density, wide measurement bandwidth, and extremely high precision.  A key feature of our system is a comb line spacing of 36 GHz, which enables direct photodetection and analysis of the resulting microwave signal in the frequency domain.  Here we present measurements of the microcomb's residual and absolute phase fluctuations.  Residual (i.e. intrinsic) noise indicates the degree that the frequencies of different comb lines are correlated, and absolute noise indicates the stability of the lines with respect to a fixed frequency reference.  Moreover, we have characterized the relationship between line spacing fluctuations and pump laser intensity noise.  This information provides a benchmark for what level of control will be required for future applications of stabilized microcombs.

To understand the intrinsic stability of the microcomb, we directly compared the line spacing at two independent portions of its spectrum.  We created two copies of the microcomb spectrum shown in Fig. \ref{lineshapes}d, using a fiber beamsplitter, and separately bandpass filtered them at 1560 nm and 1553.6 nm; see Fig. \ref{rel}a.  Independent high-speed photodetectors created electronic signals at 36 GHz.  The frequencies of these signals were reduced to baseband with a common 35.1 GHz oscillator at mixers M$_1$ and M$_2$, and the path corresponding to filter $\lambda_1$ was further shifted by 11 MHz with a low-phase-noise oscillator.  A final mixer (M$_4$) interfered the two baseband signals to reject common-mode fluctuations.  Hence, only residual phase noise of the microcomb (and noise associated with the independent paths) appeared at the output of M$_4$.  Using a commercial phase noise analyzer we recorded the single-sideband phase noise spectral density $L_\phi$, which is shown by the black line in Fig. \ref{rel}b.  For reference, the red line in Fig. \ref{rel}b shows the phase noise contribution of the 11 MHz offset signal.  The microcomb's residual phase noise is extremely low.  At carrier offsets greater than 10 kHz, $L_\phi$ is dominated by photodetection noise and would be reduced with greater microcomb optical power.  Close to carrier the residual phase noise is mostly below -100 dBc/Hz, a value comparable to the absolute phase noise of the best optical and microwave oscillators \cite{Fortier2011}.  This indicates that with appropriate frequency control such a microcomb could be harnessed for portable low-noise synthesis applications.

\begin{figure}[t!]
\centering
\includegraphics[width=0.47\textwidth]{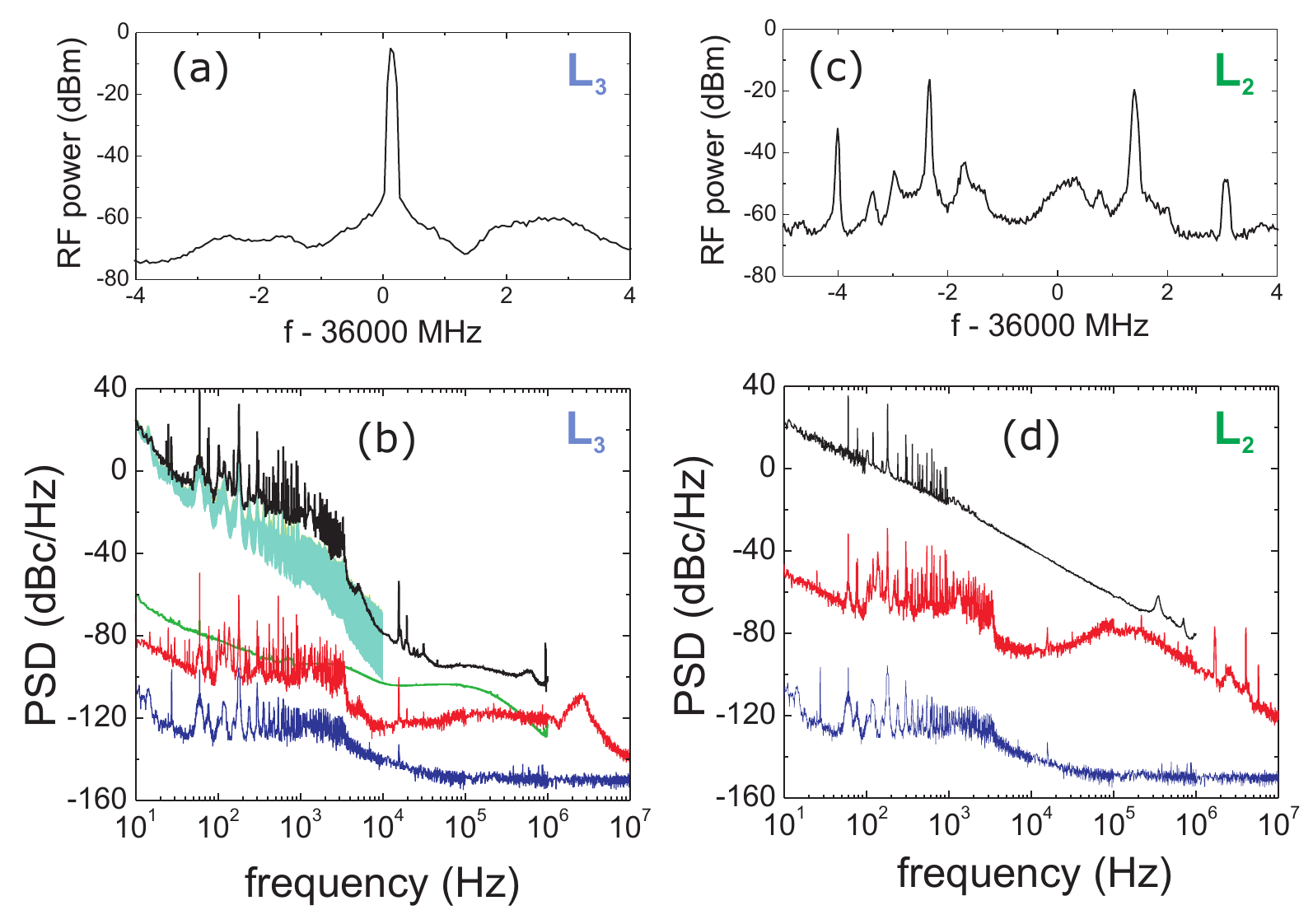}
\caption{Microcomb fluctuations measured with respect to fixed references.  For all of these experiments the entire microcomb spectrum was delivered to PD$_2$. (a)  Free-running microwave spectrum of the microcomb operating at power level L$_3$ obtained via photodetection. (b)  Power spectral density (PSD) of $L_{\phi}$ (black), line-spacing signal amplitude (red), and pump laser RIN (blue).  The RIN attains its shot noise limited value of -151 dBc/Hz beyond 100 kHz.  The cyan shaded region shows a scaling of the RIN to phase noise using the measured conversion factor $\gamma_P$ for our system.  The top range of this area corresponds to a constant $\gamma_P$ with frequency, and the bottom range accounts for the measured frequency response of pump power to line spacing.  The green line shows the phase noise contribution of the 35.5 GHz oscillator.  (c)  Microwave spectrum at L$_2$ with a 30 kHz resolution bandwidth.  Under these conditions the comb exhibits multiple line spacings near 36 GHz.  (d)  $L_{\phi}$ (black) and amplitude (red) noise of the microcomb at L$_2$.  The blue trace shows pump laser RIN.
\label{abs}}
\end{figure}

We have assessed whether the microcomb line spacing is the same at 1560 nm and 1553.6 nm by electronically counting the 11 MHz signal exiting M$_4$.  Here an offset from 11 MHz would indicate non-equidistance of the microcomb spectrum.  Previous work by Del'Haye \textit{et. al.} studied microcomb equidistance by way of an auxiliary fiber-laser frequency comb \cite{DelHaye2007}.  With access to the 36 GHz comb spacing, our measurements directly characterize microcomb equidistance for the first time.  We acquired 546 consecutive (zero dead time) one-second long measurements; Fig. \ref{rel}c shows a histogram of the frequency difference ($\Delta f$) between the microcomb and reference oscillator.  The data exhibit a Gaussian distribution with mean of -0.65 $\mu$Hz with a 25 $\mu$Hz width.  Hence, the mode spacing of the comb does not change fractionally by more than 3$\times10^{-17}$ across 6.4 nm.  The Allan deviation $\sigma_A$, shown in Fig. \ref{rel}d, demonstrates the intrinsic stability of the microcomb line spacing with $\sigma_A$=$1\times10^{-17}$ after only 400 s of averaging.

Many future applications of a microcomb will require a stable spectrum with respect to fixed frequency references, such as atomic clocks.  Here we show the microcomb line spacing's microwave spectrum, absolute phase noise, amplitude noise, and its dependence on pump laser power.  Importantly, different behavior of these was observed at power levels L$_2$ and L$_3$ that was not revealed in coarse measurements of the optical spectrum (Fig. \ref{lineshapes}c and d).  The experimental setup for these measurements was similar to Fig. \ref{rel}a, except that the beamsplitter, optical filters, and mixers M$_{1,3,4}$ were removed.  The photocurrent generated at PD$_2$ was converted to baseband with a 35.5 GHz oscillator for analysis.  For phase noise measurements we used a digital prescaler (divide-by-265) following M$_1$ to reject spurious amplitude fluctuations.  

First, we present our observations at L$_3$.  Figure \ref{abs}a shows a single microwave tone 55 dB above a broad asymmetric pedestal.  The width of this feature is limited by the resolution bandwidth of 10 kHz.   The amplitude and phase fluctuations of this tone are shown by the red and black traces in Fig. \ref{abs}b, respectively.  Notably, the absolute phase noise is between 20 and 120 dB above the residual noise, depending on the carrier offset.  For comparison the green line shows the phase noise of our commercial 35.5 GHz oscillator.  The pump laser power, with the relative intensity noise (RIN) spectrum shown by the blue trace in Fig. \ref{abs}b, significantly influences the comb line spacing.  We measured a line spacing-power dependence of $|\gamma_P|$ = 0.3 MHz/mW with a 3 dB bandwidth of $\sim$20 Hz.  The correlation between pump laser RIN and microcomb amplitude and phase noise is evident in Fig. \ref{abs}b.  Scaling the RIN by $\gamma_P$ results in the phase noise prediction shown by the cyan-shaded area.  Pump laser power fluctuations explain a significant fraction of the line spacing phase noise, and either passive or active reduction of the RIN will be crucial for future experiments.  In particular, obtaining a $f$-2$f$ heterodyne beatnote \cite{Diddams2000,Jones2000,DelHaye2009}, a key outstanding milestone in microcomb systems, will be challenging in the face of significant comb noise.

Figure \ref{abs}c and d shows a similar analysis of the microcomb line spacing's spectrum (c), amplitude (red trace in (d)) and phase fluctuations (black trace in (d)) at power level L$_2$.  Under this operating mode, the comb exhibits multiple line spacings in a few MHz range around 36 GHz.  We observe significantly larger (10's of dB) amplitude and phase noise, even though the pump laser RIN is the same as at L$_3$.  The correspondence between pump RIN and amplitude, phase noise remains evident.   We have not been able to develop a detailed understanding of, nor any strategy to mitigate, the poor performance at L$_2$.  However, these observations emphasize that measurements beyond those of the optical spectra will be critical to understanding the properties of the microcomb in all its operating regimes.

In summary, we have fabricated mm-scale, high-$Q$ optical microresonators with fused quartz.  By using them, a frequency comb with 36 GHz line spacing, 50 nm span, and $<$10 mW threshold was created near 1560 nm.  We studied the microcomb spectrum in both the time and frequency domains.  Picosecond optical pulses were generated by way of full line-by-line control of up to 15 comb lines.  Precise frequency-domain techniques enabled a direct test of comb line fractional equidistance at the 3$\times10^{-17}$ level, and a characterization of the microcomb's absolute line spacing phase noise.  Quantifying the noise of microresonator frequency combs is crucial in assessing whether they can distribute modern microwave and optical atomic standards.  In the future we will focus generating a wider comb span, and stabilizing the comb to Rb transitions at 780 nm by way of higher bandwidth comb control.

We acknowledge useful conversations with Kerry Vahala and Tobias Kippenberg, and we thank Gabe Ycas and Nathan Lemke for a thoughtful reading of the manuscript.  This work is a contribution of NIST and is not subject to copyright in the United States.


\bibliographystyle{../../../../bib_files/apsrev4-1}
\bibliography{../../../../bib_files/sp_TF,../../../../bib_files/sp_QOpt}

\end{document}